\documentclass[conference]{IEEEtran}

\usepackage{amsmath,amssymb}

\usepackage{multirow}
\usepackage[table,xcdraw]{xcolor}
\usepackage[utf8]{inputenc}
\usepackage{cleveref}
\crefname{section}{§}{§§}
\Crefname{section}{§}{§§}
\usepackage{float}
\usepackage{graphicx}
\usepackage{hyperref}
\usepackage{color}
\hypersetup{
	colorlinks   = true,
	citecolor    = blue,
	urlcolor     = blue
}
\makeatletter
\DeclareUrlCommand\ULurl@@{%
	\def\UrlLeft{\uline\bgroup}%
	\def\UrlRight{\egroup}}
\def\ULurl@#1{\hyper@linkurl{\ULurl@@{#1}}{#1}}
\DeclareRobustCommand*\ULurl{\hyper@normalise\ULurl@}
\makeatother

\usepackage{array}
\newcolumntype{L}[1]{>{\raggedright\let\newline\\\arraybackslash\hspace{0pt}}m{#1}}
\newcolumntype{C}[1]{>{\centering\let\newline\\\arraybackslash\hspace{0pt}}m{#1}}
\newcolumntype{R}[1]{>{\raggedleft\let\newline\\\arraybackslash\hspace{0pt}}m{#1}}

\DeclareFontFamily{\encodingdefault}{\ttdefault}{\hyphenchar\font=`\-}

\usepackage{caption}
\usepackage{subcaption}
\usepackage{amsfonts}
\usepackage{fixltx2e}
\usepackage{bm}
\usepackage{amsmath}
\usepackage{graphicx}
\usepackage{subcaption}
\usepackage{amssymb}
\usepackage{tikz}
\usepackage{array}
\usepackage{xcolor}
\usepackage[mathscr]{euscript}
\usepackage{mathrsfs}

\usetikzlibrary{shapes.geometric, arrows}
\tikzstyle{ADG} = [ellipse, minimum width=1cm, minimum height=.75cm,text centered, draw=black, fill=pink!30]
\tikzstyle{PDG} = [ellipse, minimum width=1cm, minimum height=.75cm,text centered, draw=black, fill=yellow!20]
\tikzstyle{SSCFP} = [ellipse, minimum width=1cm, minimum height=.75cm,text centered, draw=black, fill=green!10]
\tikzstyle{Ins} = [ellipse, minimum width=1cm, minimum height=.75cm,text centered, draw=black, fill=brown!10]
\tikzstyle{Signs} = [ellipse, minimum width=1cm, minimum height=.75cm,text centered, draw=black, fill=orange!10]

\usepackage{listings}
\lstdefinestyle{customc}{
	belowcaptionskip=1\baselineskip,
	breaklines=true,
	frame=L,
	xleftmargin=\parindent,
	language=Java,
	showstringspaces=false,
	basicstyle=\footnotesize\ttfamily,
	keywordstyle=\bfseries\color{green!40!black},
	commentstyle=\itshape\color{purple!40!black},
	identifierstyle=\color{blue},
	stringstyle=\color{orange},
}
\usepackage{pifont}
\usepackage {algorithm,algorithmicx}
\usepackage{algpseudocode}
\usepackage{footmisc}
\usepackage{footnote}
\usepackage{verbatim}

\usepackage{enumerate}
\usepackage{enumitem}
\usepackage[space]{cite}
\usepackage{footmisc}
\usepackage[nodisplayskipstretch]{setspace}

\makeatletter
\def\therule{\makebox[\algorithmicindent][l]{\hspace*{.5em}\vrule height .75\baselineskip depth .25\baselineskip}}%

\newtoks\therules
\therules={}
\def\appendto#1#2{\expandafter#1\expandafter{\the#1#2}}
\def\gobblefirst#1{
	#1\expandafter\expandafter\expandafter{\expandafter\@gobble\the#1}}%
\def\LState{\State\unskip\the\therules}
\def\pushindent{\appendto\therules\therule}%
\def\popindent{\gobblefirst\therules}%
\def\printindent{\unskip\the\therules}%
\def\printandpush{\printindent\pushindent}%
\def\popandprint{\popindent\printindent}%

\algdef{SE}[WHILE]{While}{EndWhile}[1]
{\printandpush\algorithmicwhile\ #1\ \algorithmicdo}
{\popandprint\algorithmicend\ \algorithmicwhile}%
\algdef{SE}[FOR]{For}{EndFor}[1]
{\printandpush\algorithmicfor\ #1\ \algorithmicdo}
{\popandprint\algorithmicend\ \algorithmicfor}%
\algdef{S}[FOR]{ForAll}[1]
{\printindent\algorithmicforall\ #1\ \algorithmicdo}%
\algdef{SE}[LOOP]{Loop}{EndLoop}
{\printandpush\algorithmicloop}
{\popandprint\algorithmicend\ \algorithmicloop}%
\algdef{SE}[REPEAT]{Repeat}{Until}
{\printandpush\algorithmicrepeat}[1]
{\popandprint\algorithmicuntil\ #1}%
\algdef{SE}[IF]{If}{EndIf}[1]
{\printandpush\algorithmicif\ #1\ \algorithmicthen}
{\popandprint\algorithmicend\ \algorithmicif}%
\algdef{C}[IF]{IF}{ElsIf}[1]
{\popandprint\pushindent\algorithmicelse\ \algorithmicif\ #1\ \algorithmicthen}%
\algdef{Ce}[ELSE]{IF}{Else}{EndIf}
{\popandprint\pushindent\algorithmicelse}%
\algdef{SE}[PROCEDURE]{Procedure}{EndProcedure}[2]
{\printandpush\algorithmicprocedure\ \textproc{#1}\ifthenelse{\equal{#2}{}}{}{(#2)}}%
{\popandprint\algorithmicend\ \algorithmicprocedure}%
\algdef{SE}[FUNCTION]{Function}{EndFunction}[2]
{\printandpush\algorithmicfunction\ \textproc{#1}\ifthenelse{\equal{#2}{}}{}{(#2)}}%
{\popandprint\algorithmicend\ \algorithmicfunction}%
\makeatother

\begin{document}

\title{Adaptive and Scalable Android Malware Detection through Online Learning}

\author{Annamalai~Narayanan,~Liu~Yang,~Lihui~Chen~and~Liu~Jinliang
	\\ Nanyang Technological University, Singapore.
	\\annamala002@e.ntu.edu.sg, \{yangliu, elhchen\}@ntu.edu.sg, liuj0081@e.ntu.edu.sg}
\maketitle

\begin{abstract}
It is well-known that malware constantly evolves so as to evade detection and this causes the entire malware population to be non-stationary. Contrary to this fact, prior works on machine learning based Android malware detection have assumed that the distribution of the observed malware characteristics (i.e., features) do not change over time. In this work, we address the problem of \textit{malware population drift} and propose a novel online machine learning based framework, named DroidOL to handle it and effectively detect malware. In order to perform accurate detection, the security-sensitive behaviors are captured from apps in the form of inter-procedural control-flow sub-graph features using a state-of-the-art graph kernel. In order to perform scalable detection and to adapt to the drift and evolution in malware population, an online passive-aggressive classifier is used. 

In a large-scale comparative analysis with more than 87,000 apps, DroidOL achieves 84.29\% accuracy outperforming two state-of-the-art malware techniques by more than 20\% in their typical batch learning setting and more than 3\% when they are continuously re-trained. 
Our experimental findings strongly indicate that online learning based approaches are highly suitable for real-world malware detection.
\end{abstract}

keywords | Online Learning, Graph Kernels, Malware Detection

\IEEEpeerreviewmaketitle

\section{Introduction}
\label{sec:intro}
Recently, malware detection for mobile platforms such Android has evolved as one of the challenging problems in the field of cyber-security. The number of new Android malware applications (apps for short) and their capabilities have grown tremendously in recent years. For instance, Kaspersky reports \cite{Kas} detecting 4 million malware infections in 2015 which is a 216\% increase over 2014. The sheer volume and growth rate of Android malware highlights an imperative need for developing sound and scalable automated malware detection process \cite{Drebin,DroidSift,Adagio,AppContext,CSBD,MLMalDetect,s&p}.

Malware detection using Machine Learning (ML) techniques is predominant in various platforms (such as Windows, Android and the web) for more than a decade \cite{Drebin,DroidSift,Adagio,AppContext,CSBD,MLMalDetect,s&p}. This is because, these methods automatically learn the characteristics that distinguish malware, when trained using a collection of malware and benign samples making them amenable for automated detection. ML based approaches extract features from an apps’ behaviors and apply standard ML algorithms to perform binary classification. These approaches typically use semantic features such as system calls/Application Programming Interfaces (APIs) invoked, resources and privileges used, control- and data-flows inside apps’ execution to detect malicious behavior patterns \cite{Drebin,DroidSift,Adagio,AppContext,CSBD,MLMalDetect,s&p}.

Most of these malware detection techniques are based on batch-learning classifiers. Meaning, the detection model is built using a batch of labelled benign and malware samples and is subsequently used to predict whether a given new sample is benign or malicious. These batch-learning based methods are typically plagued by two challenges that make them unsuitable for real-world large-scale malware detection: 
\begin{itemize}[leftmargin=*]
	\setlength\itemsep{0em}

\item \textbf{Population drift.} Though batch-learning based solutions are promising, their success is predicated on an important assumption that may not hold for the malware detection problem. 
That is, they assume that the malware population (i.e., training data) used to build the detection model does not change over time. However, malware does not fit this profile. The entire population of malware is constantly evolving due to various reasons such as exploiting new vulnerabilities, and evading novel detection techniques. This evolution has a profound impact on malware characteristics and thereby on malware features. This makes the collection of malware identified today not a representative of the ones generated in the future. This phenomenon is an epitome of \textit{population drift} \cite{ConDrift}. Since new malware features emerge and importances of features change over time, this population drift leads to \textit{concept drift} \cite{MalURL,ConDrift}.

\item \textbf{Volume.} As noted before, malware grows at an alarming rate and hence a scalable classifier is of paramount importance to practical large-scale malware detection. In order to keep abreast with drifting population, batch learners have to be frequently re-trained with huge volumes of data. Hence they pose severe scalability issues when used in the Android malware detection context where we have millions of samples already and thousands streaming in every day. Retraining frequently with such a volume renders them computationally impractical.
\end{itemize}

\textbf{Our Approach.} We take these two challenges into consideration and propose DroidOL, an accurate, adaptive and scalable malware detection framework based on online learning, where we continuously retrain the model upon receiving each labeled sample and make prediction of a new sample using the updated model. We demonstrate that online learning based solutions are better suited for practical large-scale automated malware detection for two reasons: 
\begin{itemize}[leftmargin=*]
	\setlength\itemsep{0em}
\item The detection model needs to adapt to changes in malware features over time, automatically. 
\item Online learning based solution can process large numbers of apps more efficiently than batch methods.
\end{itemize}

DroidOL’s achieves superior accuracy through extracting high quality features from inter-procedural control-flow graphs (ICFGs) of apps, which are known to be robust against evasion and obfuscation techniques adopted by malware \cite{AppContext,CSBD}. To this end, we use the Weisfeiler-Lehman (WL) graph kernel \cite{WLK} that supports explicit feature vector representation of graphs to extract semantic features from ICFGs.
DroidOL's adaptiveness and scalability are achieved through use of online learning. 

Hence, the primary contribution of our paper is the successful application of online learning algorithms to the problem of Android malware detection. To the best of our knowledge, we are the first to propose such a framework and demonstrate its capabilities to handle \textit{population drift}.

\textbf{Experiments.} DroidOL is evaluated through large-scale experiments on a recent real-world dataset of more than 87,000  apps.
 It is compared against two state-of-the-art  batch-learning based Android malware detection techniques. DroidOL achieves 84.29\% accuracy outperforming state-of-the-art techniques by more than 24\% in their typical batch-learning and more than 3\% when they are re-trained. Subsequently, we show that continuous retraining over newly emerging features is crucial for adapting the detection model to detect new or evolving malware. 
  
In summary, our paper’s contributions are as follows:
\begin{itemize}[leftmargin=*]
	\setlength\itemsep{0em}
\item We propose and develop DroidOL, an accurate, scalable and adaptive Android malware detection framework which is based on online learning, where we do not assume that the malware population is stationary (in \S \ref{sec:fm}).
\item We conduct and report a large-scale comparative analysis of our framework against several re-trained variants of two state-of-the-art  malware detection solutions on a sizable dataset of more than 87,000 apps (in \S \ref{sec:eval} and \ref{sec:understanding}).
\end{itemize}

The paper is organized as follows. We begin by discussing the related works on malware detection and also present our motivations in \S \ref{sec:rwm}. The DroidOL framework is introduced in \S \ref{sec:fm}. Implementation details are discussed in \S \ref{sec:impl}. DroidOL's evaluation, comparative analysis and relevant discussions are presented in sections \ref{sec:dataset}, \ref{sec:eval} and \ref{sec:understanding}. Conclusions are presented in section \ref{sec:conc}.

\section {Related Work \& Motivation}
\label{sec:rwm}
ML based approaches are popular over the past decade for malware detection. Many existing works have successfully applied ML techniques for malware detection on various platforms such as Windows, Android and the web. 
\subsection {Related Work - Android Malware Detection}
\label{ss:rw}
\subsubsection{Primitive Approaches}
In the case of Android, Crowdroid \cite{Crowd}, Drebin \cite{Drebin} and DroidAPIMiner \cite{DroidAPIMiner} are noticeable among the early approaches on ML based malware detection. These methods were designed to detect malware operating on the initial versions of Android, performing simpler attacks such as making premium-rated calls/SMS. Hence, they leveraged on primitive features such as system calls, Android APIs and permissions. These techniques detect malware through identifying suspicious usage patterns of the aforesaid features. In particular, Crowdroid \cite{Crowd} uses Linux system call sequences as features. Drebin \cite{Drebin} uses APIs, permissions, components, accessed URLs and Intent filters as features. DroidAPIMiner \cite{DroidAPIMiner} considers sensitive APIs along with parameters and package level information as features. Even though these features are good enough for detecting simpler malware, they are easily evaded by modern malware that perform sophisticated attacks \cite{DroidMiner,DroidSift}.
\subsubsection {Robust Approaches}
In order to detect stealthy malware, recent approaches leverage on two type of detection: (1) information-flow based detection and (2) graph based structural detection.

\textbf{Information-flow based detection.}
These methods track the flow of sensitive information inside the apps’ execution and detect malware by spotting suspicious flows. Even though these methods are highly precise, they exhibit poor scalability due to the expensive data-flow analysis they leverage on. Hence such methods are not suitable for practical large-scale malware detection \cite{Drebin,Adagio}.  Mudflow \cite{Mudflow}, is a prototypical example of these types of detection methods. 

\textbf{Graph based structural detection.} Graphs offer a natural way to model the sequence of activities that take place in a program. Hence they serve as amenable data-structures for detecting malware through identifying suspicious activity sequences. For this reason, graph representations such as call-graphs, control- and data-flow graphs, control-, data- and program-dependency graphs have been widely used for malware detection in conjunction with graph mining techniques \cite{DroidMiner,Adagio,AppContext,CSBD,DroidSift,s&p}. In the case of Android, DroidMiner \cite{DroidMiner} and Allix et al. \cite{CSBD} proposed to use control-flow graph based features to perform structural malware detection. DroidSIFT \cite{DroidSift} and AppContext \cite{AppContext} proposed a more robust approach by including the contextual information of security-sensitive activities (i.e., whether or not the user is aware of such an activity) along with structural information captured through graphs.

These solutions are plagued by two limitations: \\
(1) \textit{Loss of expressiveness:} These solutions follow a naïve approach to vectorise the graphs such as taking only individual nodes into consideration without their neighborhood. This leads to loosing the expressiveness of graphs.\\
 (2) \textit{Poor efficiency:} Many classic graph mining based approaches (e.g., \cite{s&p}) are NP hard and have severe scalability issues, making them impractical for real-world malware detection \cite{Adagio}.

\textbf{Graph Kernels for malware detection.} Recently, efficient and expressive graph kernels such as WL kernel \cite{WLK} have been proposed and widely adopted in many application areas (e.g, bio- and chemo-informatics, computer vision, etc.). Some of these kernels also support building explicit feature vector representations of graphs. Taking notice of such a development, two approaches, Adagio \cite{Adagio} and Sahs et al. \cite{MLMalDetect} used graph kernels to perform structural detection of Android malware.

\subsection{Motivation}
\label{ss:mot}
While all the above mentioned works focus on engineering robust features that can detect malware effectively, they do not address a key practical aspect of malware detection problem|\textit{malware evolution}.  As discussed in \S \ref{sec:intro}, many analytical studies such as \cite{Genome, ConDrift} have clearly highlighted that malware evolves in terms of its characteristics for various reasons.
The evolution inevitably leads to profound changes in the malware features over time i.e., \textit{concept drift}. This poses two challenges:\\
(C1) the detection model has to automatically adapt to the concept drift.\\
(C2) the detection model has to consider and account for new features that emerge over time.

\textbf{Malware detection as a data-stream classification problem.}
While none of the existing Android malware detection techniques address this concept drift, we get a clear motivation to do so. 
While all the existing works consider malware detection as a \textit{static classification problem} with a fixed training and test set, we note that this is against the real-world situation. Hence, we model it as a \textit{data stream classification problem}. As for the two aforementioned challenges, we address the first challenge (C1) through the use of an online learning classifier which adapts itself to the drift in characteristics of samples that stream in and for the second challenge (C2), we use a \textit{dynamic feature space}, where the number of features keeps growing over time (see \S \ref{ss:tr} for a detailed explanation).

While we note that there a few approaches in the past that have adopted online learning to tackle concept drift in malware for other platforms such as Windows \cite{OLMal}, we are the first to do so for Android.

\textbf{Features.} As discussed in the previous subsection \ref{ss:rw}, graph based features have been proven effective and robust to several evasion strategies followed by malware and hence we use them. While choosing a particular graph representation, we note that building data-flow and data-dependency graphs is not scalable and call-graphs are too coarse-grained to capture the program semantics perfectly. Hence, we use ICFG representation as they efficiently capture finer program semantics.

\textbf{Graph Kernel.} Since WL graph kernel \cite{WLK} is the current state-of-the-art graph kernel, known for its expressiveness and efficiency, we use it to extract semantic features from ICFGs.
\section{DroidOL - Framework Overview}
\label{sec:fm}

\begin{figure}[t]
	\includegraphics[height=3.4cm,width=9cm]{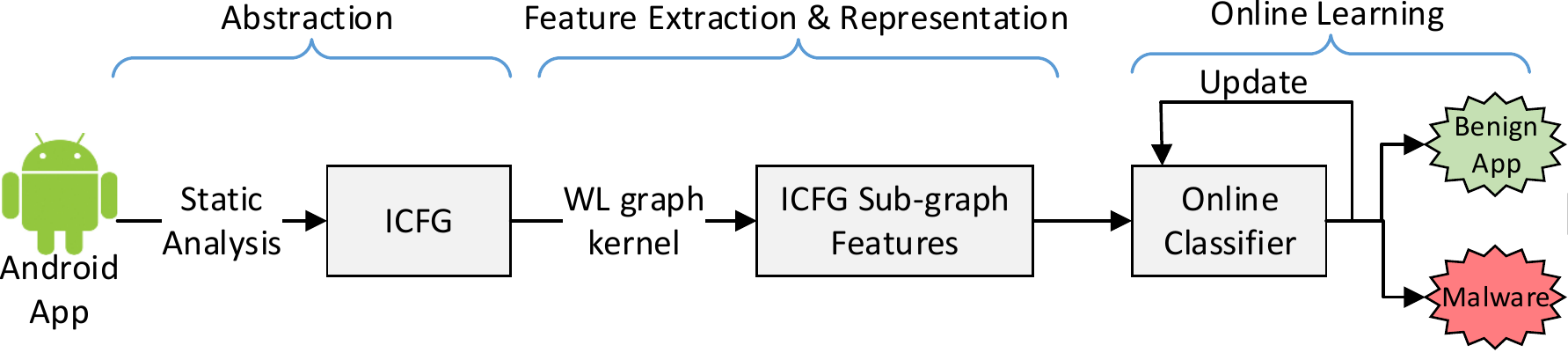}
	\caption{DroidOL framework for performing online learning based Android malware detection
	\label {fig:fm}}
\end{figure}

\begin{algorithm}[t]
	\small
	\caption{WL kernel | extracting vocabulary for feature vector representation of ICFGs}
	\label{algo:wlk}
	\textbf{Input}:\\
	$\mathcal{G} = \{ICFG_1,ICFG_2,...,ICFG_K\}$: A set of $K$ ICFGs (one for each of the $K$ apps in the dataset.)  \\
	$h$: Degree of neighborhood to be considered for label enrichment\\
	\textbf{Output}:\\
	$\Sigma$: Vocabulary of sub-graph features present across all graphs in $ \mathcal{G} $ 
	\begin{algorithmic}[1]
		
		\Procedure{Extract Vocab}{$\mathcal{G},h$}
		\LState $\Sigma \leftarrow \phi$ 
		\For {$ ICFG = (N,E,\lambda_0)\in \mathcal{G} $}
		\For {$i = $ 0 to $h$}
		\For { $n \in N$}
		\If {$i = 0$}
		\LState $\lambda_0(n) \leftarrow \lambda(n) $ 
		\Else
		\LState $\mathcal{N}(n) \leftarrow \{m\ |\ (n,m) \in E\}$
		\LState $M_i(n) \leftarrow \{\lambda_{i-1}(m)\ |\ m \in \mathcal{N}(n) \}$ 
		\LState $\lambda_i(n) \leftarrow  \lambda_{i-1}(n) \oplus sort(M_i(n))$  
		\EndIf
		\LState $\Sigma \leftarrow \Sigma \cup \lambda_i(n)$  
		\EndFor
		
		\EndFor
		\EndFor
		\LState \textbf{return } $ \Sigma $
		\EndProcedure
		
	\end{algorithmic}
\end{algorithm}

The overview of DroidOL framework which performs accurate, adaptive and scalable malware detection using online learning is presented in Fig. \ref{fig:fm}. DroidOL has three phases. We begin by performing static analysis on a given set of apps to get their ICFG representations. Subsequently, ICFG subgraph features are extracted using the WL kernel and the apps are represented as feature vectors. Finally, an online PA classifier is trained with these vectors to detect malware. Each of the phases is described in detail below.
\begin{enumerate}[leftmargin=*]
	\setlength\itemsep{0em}
	
	\item \textbf{Abstraction.}
	Our malware detection approach considers node-labeled ICFG sub-graphs as features. To extract these features, we first perform Android-specific static analysis to transform all the apps into their corresponding ICFG representations. Subsequently, the ICFG nodes are labeled with security sensitive APIs\footnote{PScout \cite{PScout}, an existing research work identifies and lists the security-sensitive Android APIs. These APIs are used for labeling our ICFG nodes.} that they access. Formally,
	$ICFG = (N, E, \lambda)$ is a directed graph where $N$ is a set of nodes and each node $n \in N$ denotes an instruction in the disassembled format. $E \subseteq (N\times N) $ is a set of edges and each edge $e(n_1, n_2) \in E$ denotes the control-flow from  $n_1$ to $n_2$. $\lambda$ is the set of security-sensitive Android APIs and $\ell: N \rightarrow \lambda$, is a labeling function which assigns an API as label to each node.
	
	\item \textbf{Feature Extraction \& Representation.}
	Once the ICFGs are constructed, (rooted) sub-graphs in these ICFGs that represent the security-sensitive behaviors in every app are extracted using the WL graph kernel \cite{WLK}. Subsequently, ICFGs are represented as feature vectors.
	 
\textit{WL kernel.} The WL kernel works by augmenting the labels of every node $ n $ with its neighborhood (up to a certain degree) in a given graph, $ G $. The frequency of these enriched node labels which denote sub-graphs around every node in $ G $ are used as features to facilitate an explicit feature vector representation of $ G$. 
	 	
	Thus the process of obtaining feature vector representation of ICFGs in our dataset using WL kernel involves two steps: (1) Building a vocabulary, $ \Sigma $ of sub-graph features present across all ICFGs, (2) Transforming every ICFG into a feature vector with $ |\Sigma|$ dimensions. Step (1) is presented in detail in Algorithm \ref{algo:wlk}.\\
	\textit{Algorithm Explanation.}
	The inputs to Algorithm \ref{algo:wlk} are $ \mathcal{G} $, a set of $ K $ ICFGs (one for each of the $K$ apps in the dataset) and $h$, the degree of sub-graphs to be considered for feature extraction. The output is a vocabulary set $ \Sigma $ that contains the unique sub-graphs upto degree $ h $ present across all ICFGs in $ \mathcal{G} $. 
	
	In each iteration $ i $ of the algorithm (lines 4 to 15), the neighborhood up to degree $ i $ around a node $ n $ is captured and condensed in the form of a neighborhood label, $ \lambda_i(n) $. These neighborhood label is what we refer to as sub-graph features. For every $ICFG_k \in \mathcal{G} $, the following process is adopted to obtain these neighborhood labels. 
	
	For the initial iteration $ i=0 $ no neighborhood information needs to be considered. Hence the neighborhood label $ \lambda_0 (n) $ is same as the original node label $ \lambda (n) $ (line 7). For $ i\!>\!0 $, the following procedure is used for label enrichment: Firstly, for a node $n \in N$, all of its neighboring nodes are obtained and stored in $\mathcal{N}(n)$ (line 9). For each node $m \in \mathcal{N}(n)$ the neighborhood label up to degree $i-1$ is obtained and stored in multiset $M_i(n)$ (line 10). Subsequently, $\lambda_{i-1}(n)$, neighborhood label of $n$ till degree $i\!-\!1$ is concatenated to the sorted value of $M_i(n)$ to obtain the current neighborhood label,  $\lambda_i(n)$ (line 11). Finally, the neighborhood label $\lambda_i(n)$ representing the $ i^{th} $ degree neighborhood around $ n $ is added to the vocabulary of sub-graph features $ \Sigma $ (line 13).

	Once the vocabulary $ \Sigma $ is obtained, we transform every $ ICFG \in \mathcal{G} $ into its corresponding vector representation by counting the frequency of every feature from $ \Sigma $ in $ ICFG_k $. 
	This procedure falls under the well-known Bag-of-Features (BoF) representation model \cite{WLK}, where every ICFG is considered as a bag of sub-graphs. 
	
	\item 	\textbf{Online Learning.}
	Once the feature vectors of all the apps in the training-set are built, we train an online PA classifier with these vectors to detect malware. PA classifier's training and update procedures are as explained below with relevant notations. 
	
	Denote the features of an app (both benign and malware) as a vector $x = [x^{(1)}, x^{(2)},...,x^{(d)}]^{T}$, 
	and its label as $y \in \{-1, +1\}$, where $-1$ indicates benign and $+1$ indicates malicious apps. The PA classifier receives a number of samples, $x_i$, and their labels, $y_i$, and trains using this labeled data. Given a new unseen sample, $x$, the goal of PA classifier is to predict the label, $y$, of this new sample based on its trained model. 
	
	PA being a linear classifier fits a linear decision boundary (i.e., hyperplane) between the positive and negative class samples. That is, the model is a weight vector, $w = [w^{(1)}, w^{(2)},...,w^{(d)}]^{T}$ which indicates the weight (i.e. relative importance) of each of the features used to predict the output label $y$. The predicted label, $\hat y$, is the sign of the inner product between $x$ and $w$:
	\begin{equation}
	\hat y = sign (x \cdot w)
	\end{equation}
	
	PA incrementally build the models in rounds. In round $t$, PA receives a sample, $x_t$ and predicts its label $\hat {y_t}$ using the current model; it then receives $y_t$, the true label of $x_t$ and updates its model based on the sample-label pair: $(x_t, y_t)$, if it makes a wrong prediction. 
	That is, $w$ is updated if the predicted label, $\hat {y_t}$ and the true label, $y_t$ of the sample $x_t$ are not the same. The goal of the PA algorithm is to update the model $w$ as minimal as possible to correct for any mistakes it commits. PA solves the following optimization with each given sample: 
	\vspace{-2mm}
	\begin{equation}
	w_{t+1} \leftarrow \operatornamewithlimits{argmin}\limits_{w} \frac{1}{2} || w_t - w ||^2
	\end{equation}
	\vspace{-6mm}
	\begin{center}
		subject to $y_i (w \cdot x_t) \geq 1$
	\end{center}
	Updates occur only when  $y_t(w_t \cdot x_t) < 1$., The closed-form update for all samples is as follows:
	\begin{equation}
	w_{t+1} \leftarrow w_t + \alpha_ty_tx_t
	\end{equation}
	where $\alpha_t = max\{\frac{1 - y_t(w_t \cdot x_t)} {||x_t||^2} , 0\}$ (we refer the reader to the original work at \cite{PA} for this derivation and further details on PA algorithm). 
	
	Once the PA classifier in DroidOL is trained with all these samples it is ready to perform malware detection at scale. It is important to note that since DroidOL is trained in an online fashion, it performs malware detection and simultaneously adapts to the changing trends in malware features by retraining on every sample it misclassifies.
	
	\textbf{Alternatives.} As practitioners involved in building an online malware detection framework, we do not have any vested interest in a particular algorithm for online learning. Ultimately, we wish to determine the algorithm that scales well to problems of our size and yields the best performance. 
	To that end, we experimented with other well-known online learning algorithms such as Online Perceptron (OP) and Stochastic Gradient Descent learning based Logistic Regression (LR\textsubscript{SGD}) along with PA. Since PA offered the best results in terms of accuracy, it is used in DroidOL. We believe this is because OP and LR\textsubscript{SGD} do not imbibe the notion of classification confidence and treat all misclassifications equally, unlike PA. PA, on the other hand, updates more aggressively when the margin of error is large and less aggressively in when it is small. While we note that evolving classifiers such as pClass \cite{MPratama} could be used for handling concept drift, we prefer PA over such methods, as it offers better efficiency.	 
	
	
	
\end{enumerate}

\section{Implementation \& Comparative Analysis}
\label{sec:impl}
We implemented the DroidOL framework in approximately 15,600 lines of Python and Java code. Soot\footnote{{http://sable.github.io/soot}}, a popular Android static analysis workbench is used for constructing the ICFGs of the apps in our dataset. Scikit-learn\footnote{{http://scikit-learn.org}} toolbox is used for all our ML functionalities.

We compare our online learning based detection against two state-of-the-art Android malware detection solutions, namely Drebin \cite{Drebin} and Allix et. al.’s \cite{CSBD}. It is noted that both these methods use batch learning classifiers.

\textbf{Drebin} \cite{Drebin} is well-known for its scalable and explainable detection. It extracts light-weight features such as APIs, permissions, URLs accessed, names of components from apps and subsequently, trains a linear SVM classifier to distinguish malware from benign apps.

\textbf{Allix et al.} \cite{CSBD} recently proposed another scalable approach using structural features, namely CFG signatures. Therefore, we refer to this technique as CFG-Signature Based Detection (CSBD) in the reminder of the paper. 
CSBD constructs CFGs of individual methods and encodes them as text-signatures. Subsequently, a Random Forest classifier is trained with these signatures to detect malware.

\section{Data collection}
\label{sec:dataset}
\begin{table}[t]
	\centering
	\small
	\caption{Dataset with apps dated from Jan 1, 2014 to Aug 13, 2014}
	\label{tab:ds}
	\begin{tabular}{|l|c|c|}
		\hline
		{\bf Market Name} & {\bf \# of Benign Apps} & {\bf \# of Malware} \\ \hline
		Google Play  & 39156                   & 26178                \\ \hline
		Anzhi         & 2957                    & 12260                \\ \hline
		AppChina      & 1845                     & 4154                \\ \hline
		SlideMe       & 289                     & 132                  \\ \hline
		HiApk        & 65                      & 157                  \\ \hline
		FDroid       & 29                      & 2                   \\ \hline
		Angeeks      & 6                       & 27                  \\ \hline
	\end{tabular}
	
\end{table}
\begin{figure}
	\includegraphics[height=3.7cm, width=8.7cm]{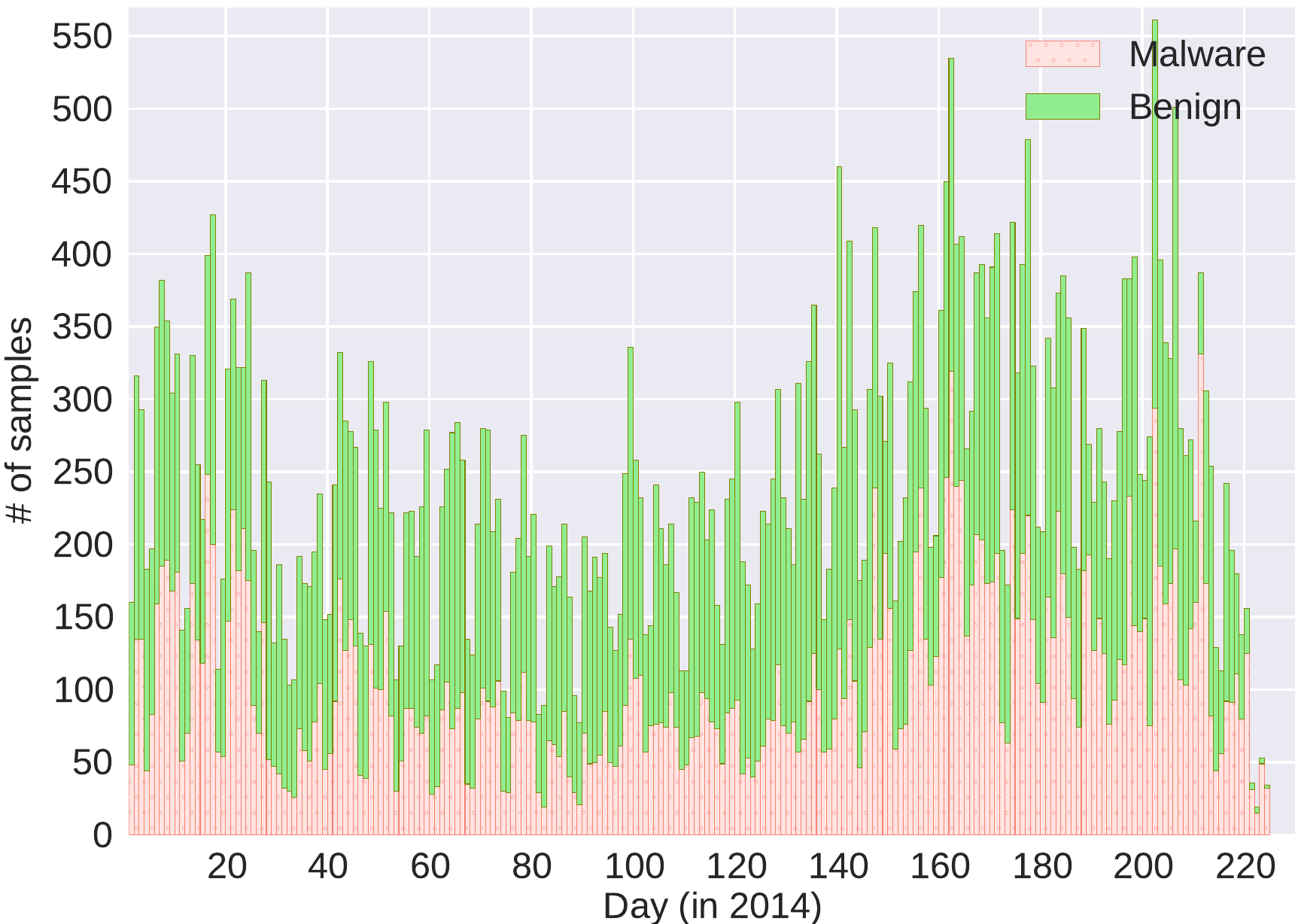}
	\caption{ Timeline based distribution of apps in our large-scale dataset (with malware and benign apps proportions).
		\label {fig:ds}}
\end{figure}
We evaluate our approach on a recent large-scale real-world dataset of 87,257 apps collected in-the-wild. These apps are collected from seven different Android markets\footnote{Google Play:  {https://play.google.com/store}, Anzhi: {www.anzhi.com}, AppChina: {{www.appchina.com}}, SlideMe: {{www.SlideME.org}}, HiApk: {{www.hiapk.com}}, FDroid: {{www.fdroid.org}} and Angeeks: {{http://apk.angeeks.com}}}, namely, \textit{Google Play, Anzhi, AppChina, SlideMe, HiApk, FDroid} and \textit{Angeeks}, in 2014.  
Following the common practice in software security research, we use the \textit{Virus Total}\footnote{{https://www.virustotal.com/}} web portal which hosts malware detection services from more than 40 Anti-virus scanners to determine the ground-truth labels of the apps. We infer that dataset contained 44,347 benign and 42,910 malware apps.
The composition of the dataset is presented in Table \ref{tab:ds}. It is noted that this is a subset of a large collection of apps used in \cite{KevinThesis}.

The date of creation of these apps fall in a span of 224 days starting from 1 Jan’14 to 13 Aug’14. 
We intend to sort these apps according to their date of creation and emulate a live feed of apps to the malware detectors considered in our experiments to examine how they cope up with the drift in the malware characteristics over time. To this end, we divide all the apps in the dataset (both benign and malware) into batches according to their date of creation. Hence we have 224 batches, one for each day. 
The resulting time-line based distribution of the two datasets are presented in Fig. \ref{fig:ds}.



 \begin{figure*}[ht!]
 	\centering
 	\includegraphics[height=6cm,width=18cm]{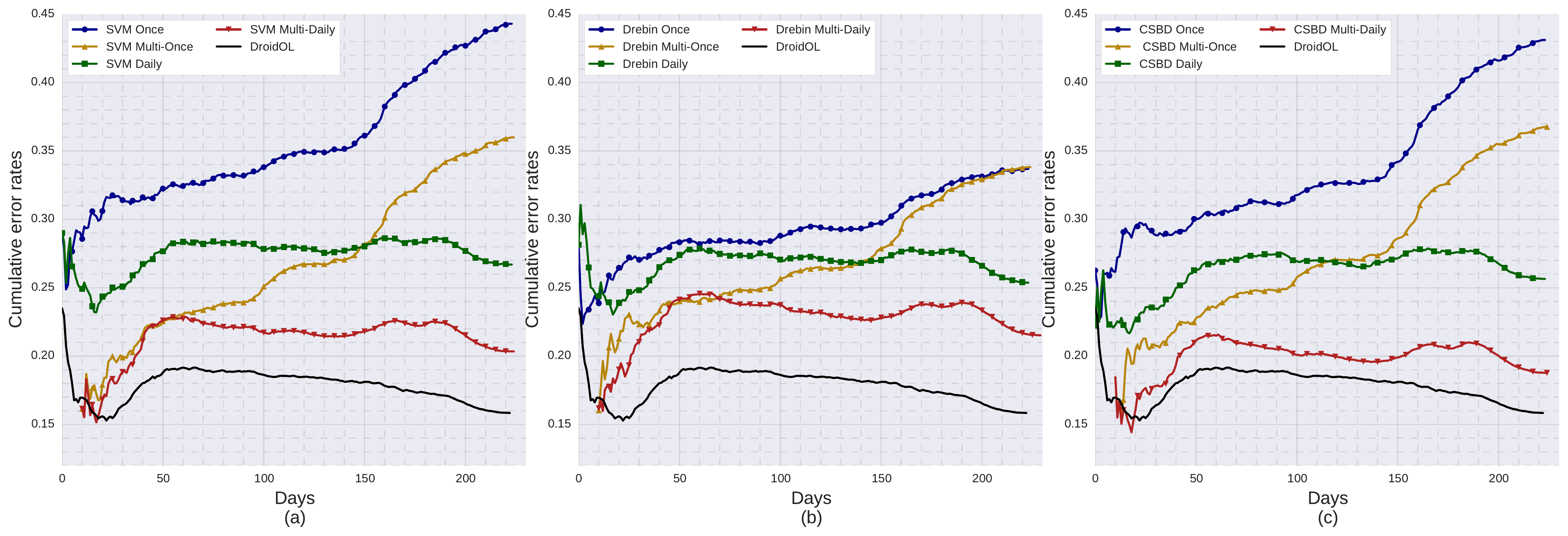}
 	\caption{Cumulative error rates for DroidOL vs. batch learning algorithms (a)  DroidOL Vs. four variants of  SVM (b) DroidOL Vs. four variants of  Drebin (c) DroidOL Vs. four variants of  CSBD.
 		\label {fig:bvso}}
 \end{figure*}
\section{Evaluation}
\label{sec:eval}
In this section, we evaluate DroidOL's  accuracy and adaptiveness using the emulated live feed of apps. To this end, we address the following questions:\\ 
(1) Does DroidOL's online learning provide any benefit over batch learning for malware detection?\\ 
(2) How does DroidOL's accuracy compare to light-weight state-of-the-art malware detection techniques?\\
(3) Is there a particular training regimen that fully realizes the potential of online classifier? 


\subsection{Advantages of Online Learning}
\label{ss:advol}

We start by evaluating the benefit of using online over batch learning for the problem of malware detection in terms of detection accuracy | in particular, whether the benefit of DroidOL's efficiency comes at the expense of accuracy. Specifically, we compare DroidOL's PA based classification against four different training configurations of a canonical batch learning classifier. We used Linear SVM as our canonical batch learner. 
It is noted that evaluations with other batch algorithms such as logistic regression yielded similar results. 

\textbf{Batch Learning Configurations.} For SVM, we experiment with the following variants: SVM-Once, SVM-Daily, SVM-MultiOnce, and SVM-MultiDaily. For SVM-Once,  
SVM classifier is trained only once on the apps from Day 1 (1 Jan’14). This model is tested on all other days without retaining. For SVM-Daily, the classifier is retrained after every day; however, only one previous day’s samples are used for every retraining — e.g., 11 Jan’14 results reflect training on the apps created on 10 Jan’14, and testing on 11 Jan’14 apps. SVM-MultiOnce is similar to SVM-Once and SVM-MultiDaily is similar to SVM-Daily, however, with the size of the batch for training and retraining covers 10 days instead of 1. 
In summary, for Once and MultiOnce variants, the model is never re-trained and for Daily and Multi-Daily, the model is re-trained in a sliding window fashion over the batches of data. The size of MultiDaily training sets is determined to be 10-day batches based on the data that our evaluation machine with 32 GB RAM can handle.

Figs. \ref{fig:bvso} (a) shows the cumulative error rates of DroidOL in comparison to the aforementioned variants of SVM. The following observations are made from Fig. \ref{fig:bvso} (a):
\begin{itemize}
\item Updating the detection models over time is essential to detect new malware as shown by SVM-MultiDaily and SVM-Daily outperforming SVM-MultiOnce and SVM-Once, respectively. 

\item Training on significantly more data improves the performance, as illustrated by SVM-MultiDaily  and SVM-MultiOnce outperforming SVM-Daily and SVM-Once, respectively. However, it is noted that there is a fundamental limit on the amount of data a batch-learning technique could train on because of the storage, memory and time requirements.

\item Third, DroidOL consistently outperforms all the batch learnt variants. In particular, it outperforms the best re-trained variant of SVM by more than 4\% cumulative error rate. This is because DroidOL is able to adapt to the changes in the malware characteristics instantaneously as well as retain significantly useful information from the past.
\end{itemize}

\subsection{Comparison with state-of-the-art Malware Detectors}
\label{ss:compsoa}
The above-mentioned SVM variants use the same features as DroidOL and hence it is sufficient to compare online and batch learning paradigms. However, this does not reflect the significance of DroidOL as a practical malware detector in the context of current state-of-the-art malware detection techniques. In order to study this, we compare DroidOL with two state-of-the-art malware detectors, namely, Drebin \cite{Drebin} and CSBD \cite{CSBD}. 

For this comparison, we follow the same batch learning configurations described in \S \ref{ss:advol} to arrive at the four variants of these  techniques: Drebin/CSBD-Once, Drebin/CSBD-Daily, Drebin/CSBD-MultiOnce and Drebin/CSBD-MultiDaily.

The results of this comparison are presented in Fig \ref{fig:bvso} (b) and (c). From these figures we make the following inferences: 
\begin{itemize}
\item For both these methods, the trends in performance of all four variants are similar to those of the SVM variants. Hence the observations made in \S\ref{ss:advol} on frequently updating the models and training with more data, hold. The error rates of best-performing variants of these methods are comparable to that of SVM-MultiDaily.

\item DroidOL consistently outperforms the best performing variants of both the methods. Particularly, it outperforms Drebin-MultiDaily by more than 5\% and CSBD-MultiDaily by more than 3\%. This reaffirms the suitability of online learning for practical large-scale malware detection.
\end{itemize}

\begin{center}
	\begin{figure}[t]
		\centering
		\includegraphics[height=4cm,width=7cm]{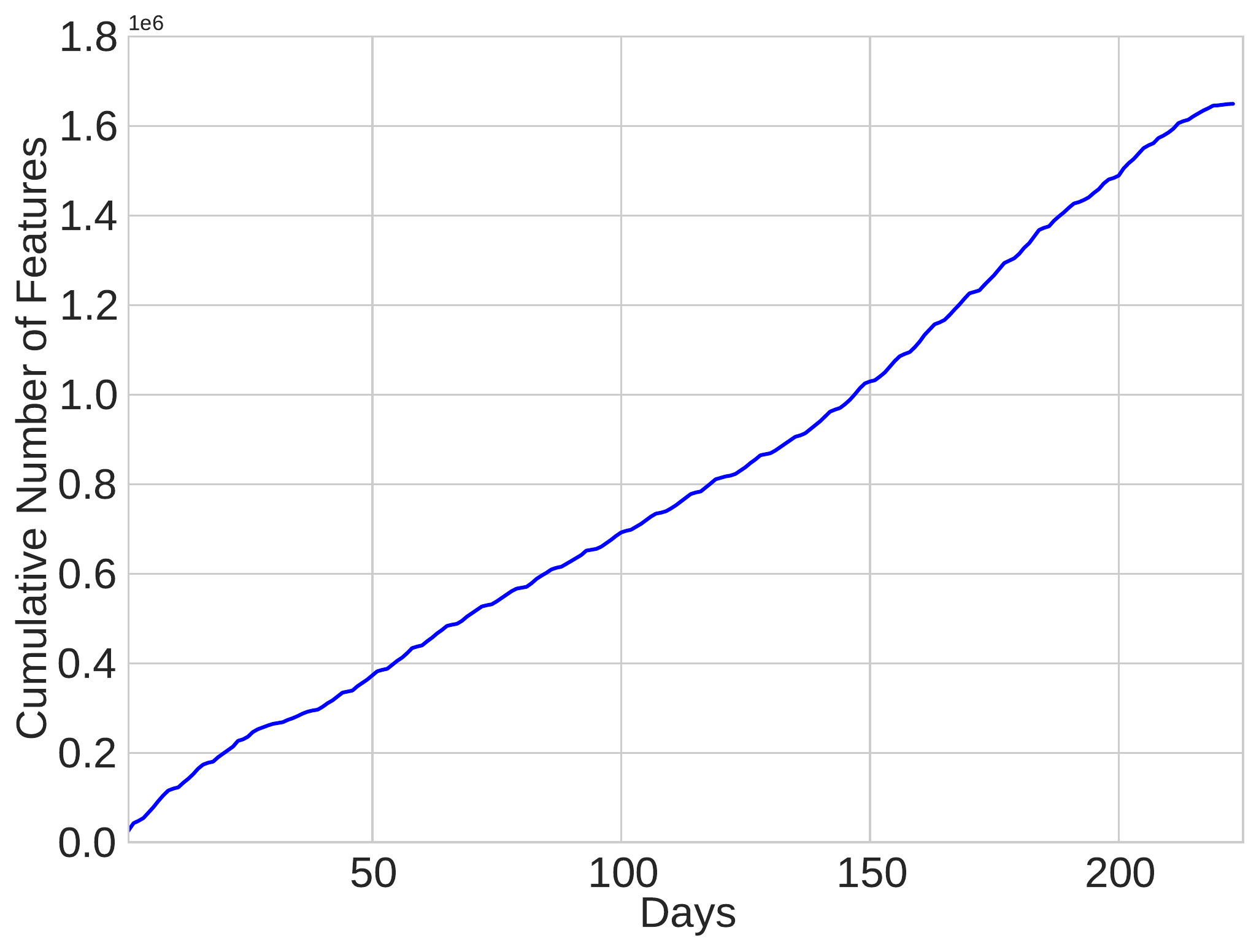}
		\caption{Cumulative number of features observed over time for
			our large-scale dataset.
			\label {fig:feat_growth}}
	\end{figure}
\end{center}

\begin{figure}[t]
	\centering
	\includegraphics[height=4.2cm,width=8cm]{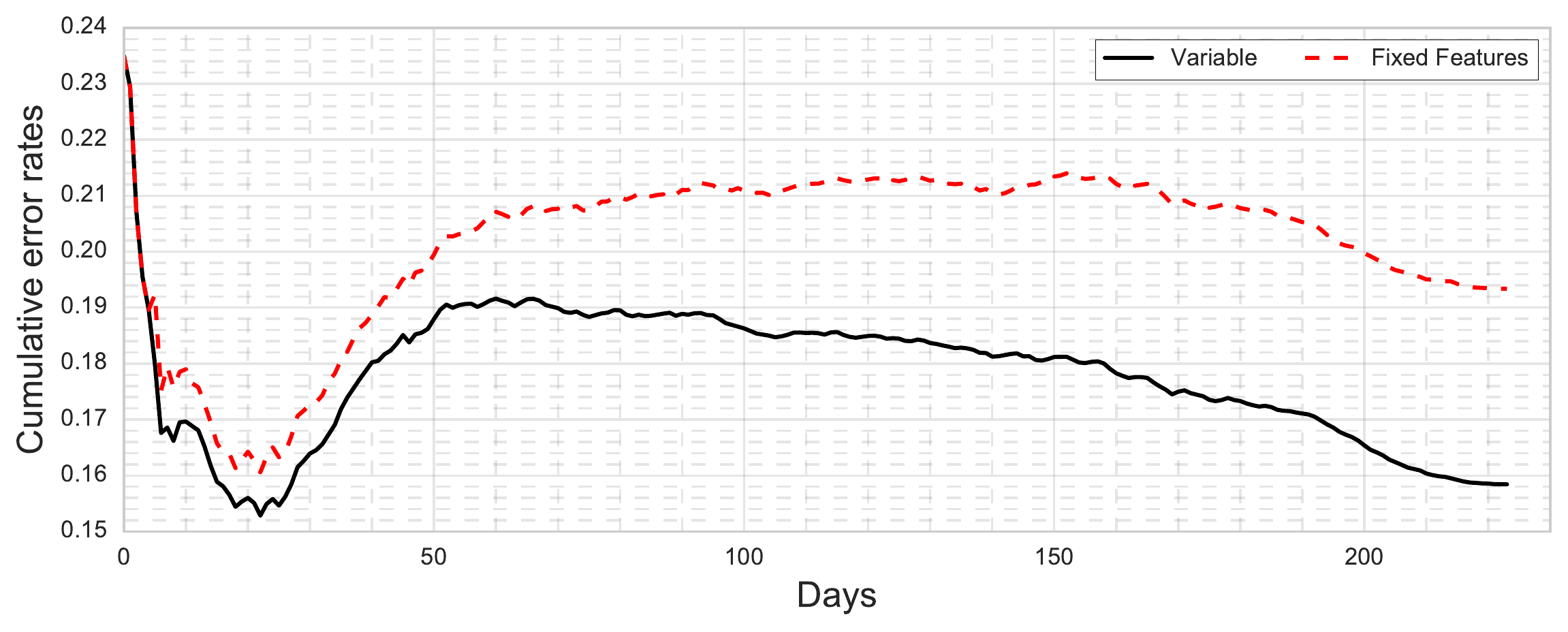}
	\caption{  Benefits of using variable-feature sets
		over fixed-feature sets.
		\label {fig:var_fixed}}
\end{figure}

\subsection{Training Regimen}
\label{ss:tr}
Since DroidOL’s feature extraction using WL kernel is based on BoF model, our number of features grows as the samples stream in. Fig. \ref{fig:feat_growth} shows the cumulative number of features for each day of the evaluations in our dataset, representing the feature space growth. Each day’s total includes new features introduced that day and all the old features from previous days. We obtained a total of 13,655 features from the malware and benign samples encountered on Day 1 (1 Jan'14). The dimensionality grows quickly as we extract new sub-graph features from samples encountered every day and while reaching the final day (13 Aug'14), we have accumulated 1,653,496 features.
This phenomenon of feature space growth is common across many techniques including Drebin and CSBD. This is because Android apps evolve over time for various reasons such as capability enhancements, bug fixes, using newly introduced Android functionalities and adapting to changes in Android framework APIs \cite{AppContext,DroidSift}. This evolution results in newly observed characteristics which translate into new features from a ML view point.

Now, we are posed with the question: \textit{should we consider these new features that emerge every day?} To address this, we devise two types of \textit{training regimen}, namely, \textit{variable feature-set training} and \textit{fixed feature-set training}.\\
\textbf{Fixed feature-set training regimen.} Under the \textit{fixed feature-set} regimen, we train using a pre-determined set of features for all evaluation days. That is, we fix the features to those encountered up to Day 1, then we use those 13,655 features for the whole experiment.\\
\textbf{Variable feature-set training regimen.} Under the \textit{variable feature-set} regimen, we allow the dimensionality of our PA classifier to grow with the number of new features encountered; on the last day, for instance, we classify with more than 1.6 million features. Implicitly, examples that were introduced before a feature \textit{i} was first encountered will have value 0 for feature \textit{i}. 

Fig. \ref{fig:var_fixed} shows the importance of using \textit{variable feature-set training} over \textit{fixed feature-set training}. We see that the performance for fixed feature regimen is significantly and consistently inferior to the variable feature regimen. The latter regimen has a 3.3\% lesser cumulative error rate. This reveals that, continuous retraining with a \textit{variable feature-set} allows a model to successfully adapt to new data and new features on a sub-day granularity. This adaptiveness is critical to realize the full benefits of online learning. This indicates that choosing the right training regimen can be just as important as choosing the right classifier. The aforementioned training regimens can help online algorithms stay abreast of changing trends in malware and benign apps’ features.

\section{Understanding the performance}
\label{sec:understanding}
In this section we seek to get deeper insights into the superior performances of DroidOL's online learning and to understand how characteristics of the Android malware detection task affects its performances. Specifically, we evaluate and quantify the importance of long term memory and fast model update.
To this end, we pose the following question: \textit{Why does DroidOL outperform batch-learning solutions in detecting malware, even when they 	are re-trained?}

\textbf{Dataset.} 
In order to provide insights into our results, we need the notion of similarity among the malware samples. More specifically, we need the malware to be grouped according to their families. Malware familial analysis and grouping is currently a manual process \cite{Drebin,AppContext}. It is impractical to group apps from our large-scale dataset according to their families. Hence we use a well-known benchmark dataset that has malware readily grouped according to their family, namely, Drebin\textsubscript{5k}\footnote{In order to distinguish the dataset provided by Drebin authors from the Drebin malware detection technique discussed in \S \ref{sec:eval}, we refer to the dataset as Drebin\textsubscript{5K}.} \cite{Drebin}. Drebin\textsubscript{5K} contains a total of 5560 malicious Android apps belonging to 179 malware families. The date of creation of these samples lie in the range: Mar'09 to Oct'12. 

\textbf{Familial similarity and Notations.} We now introduce the notion of familial similarity between a pair of malicious apps. We consider two malware $m_1$ and $m_2$ as variants, denoted by $m_1\sim m_2$, if they belong to the same malware family. For instance, Drebin\textsubscript{5K} dataset contains 965 samples belonging to the \textit{FakeInstaller} family. We consider each of pair of them as variants. This stems from the fact that malware belonging to the same family exhibit similar malicious characteristics, making them exhibit similar semantic features from a ML view point.

Subsequently, for each malware $m$, set $M$ contains all the malware belonging to the same family as $m$. We calculate the time difference between the creation of $m$ and every other malware, $m' \in M$, denoted as $\triangle (m,m')$. We call the minimum delay between another variant of the same family as $ m $ as \textit{minimum variant delay}, denoted by $\triangle_{min}(m)$. Similarly, we call the maximum delay between another variant of the same family as $ m $ as \textit{maximum variant delay}, denoted by $\triangle_{max}(m)$.

If there are no variants of for a malware $m$, its default values for $\triangle_{min}(m)$ and $\triangle_{max}(m)$
are 0 and $D$, respectively, where $D$ is the difference between the date of creation of $m$ and that of the latest app in Drebin\textsubscript{5K} dataset.

\begin{figure}
	\centering
	\includegraphics[height=3cm,width=8cm]{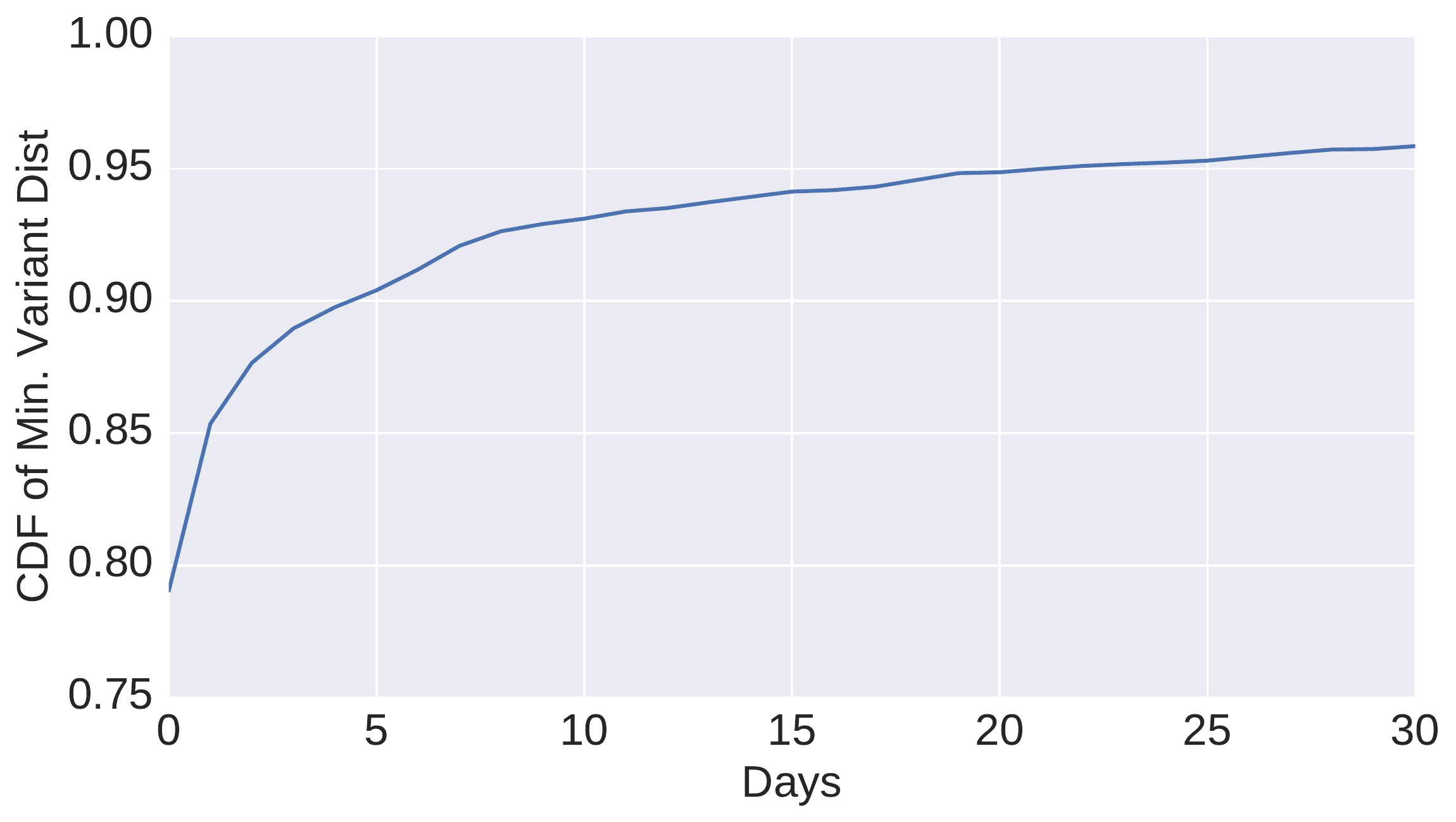}
	\caption{ The Head of CDF of minimum variant distance in Drebin\textsubscript{5K} dataset
		\label {fig:cdf}}
\end{figure}

\begin{figure}
	\centering
	\includegraphics[height=3cm,width=8cm]{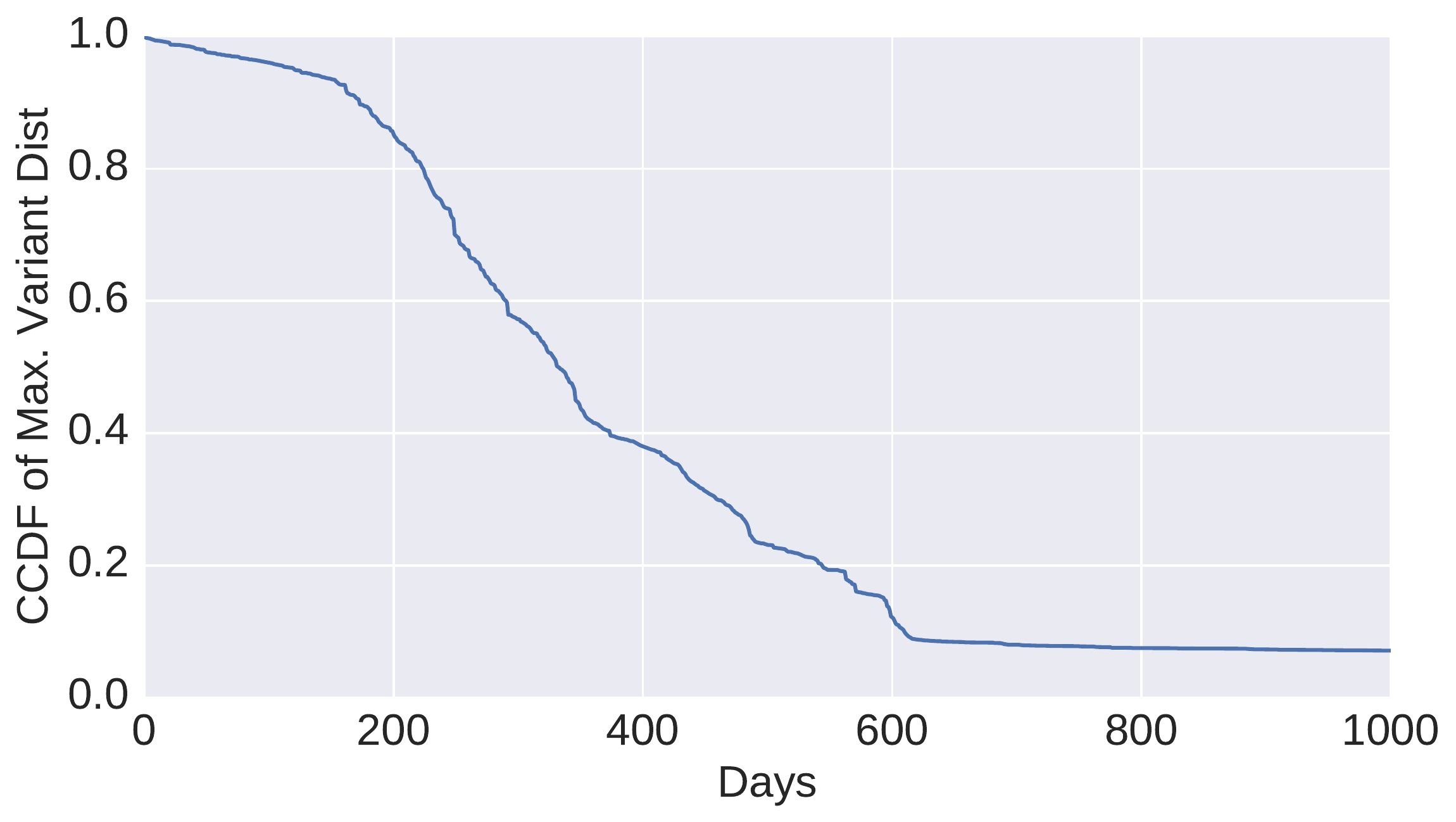}
	\caption{ The Head CCDF of maximum variant distance in Drebin\textsubscript{5K} dataset
		\label {fig:ccdf}}
\end{figure}
\textbf{Importance of fast model update.}
Fig. \ref{fig:cdf} depicts the head of the cumulative distribution function (CDF) of the \textit{minimum variant delay} of the malware samples in the Drebin\textsubscript{5K} dataset. It is clear that nearly 80\% malware have a \textit{minimum variant delay} of zero days. Meaning, for a given malware sample, at least one other variant of the same of family is built (and probably released) on the same day itself. This is understandable as malware authors aim to maximize their gains by launching several polymorphic variants of their malware, around the same time, leveraging on techniques such as obfuscation. Hence in order to keep abreast with quick-succession releases batch-learning solutions have to be updated at least daily (ideally, they have to be updated continuously). This explains why re-training SVM/Drebin/CSBD solutions every day yields higher classification accuracy in sections \ref{ss:advol} and \ref{ss:compsoa}. Therefore, unless the detector is updated continuously to reflect the most recent features present in the last malware, it will not be able to effectively detect a majority of its variants.

\textbf{Importance of long term memory.}
Fig. \ref{fig:ccdf} depicts the head of the complementary cumulative distribution function (CCDF) of the \textit{maximum variant delay} of all the malware samples in the Drebin\textsubscript{5K} dataset. We observe that more than 40\% of malware have a \textit{maximum variant delay} of more than a year. Which means, a significant number of malware families keep evolving for a few years. This is understandable as malware authors make new variants either to evade known detection techniques or to improve/enhance their attacks for a prolonged period of time. 

Therefore, if the batch size is limited to a few days or months, these 40\% of malware would not have many similar variants in the batch. This leads to a significant reduction the detection accuracy because the detection model, has not yet learned enough on malware that is similar to these malware by the time it needs to classify them. 
In general, Fig. \ref{fig:ccdf} demonstrates that classifying malware require long term memory. This explains why extending SVM/Drebin/CSBD batch sizes yielded better classification performance in sections \ref{ss:advol} and \ref{ss:compsoa}. However, when using batch-learning based detection solutions, the batch size is limited by the amount of available memory. Particularly, for a problem such as Android malware detection where we have millions of samples in a year and typical ML solutions extracting thousands of features, having a batch size of more than a year is not practical.  On the other hand, online learning algorithms do not have this limitation. They retain significant useful information from all of the malware that they have seen. In other words, online algorithms operate with effectively infinite batch size. This explains why they have an edge over batch-learning solutions in our experiments.

\textbf{Summary. }From figures \ref{fig:cdf} and \ref{fig:ccdf} it is clear that to perform reasonably accurate and adaptive malware detection, the batch-learning solutions should re-train every day with a batch size of at least a year. This strategy is highly expensive in terms of computation time and resources, rendering it impractical. Hence, we can confidently conclude, online learning based methods which do not have such computational limitations and inherently adaptive are better suited for Android malware detection.


\section{Conclusions}
\label{sec:conc}
In this paper, we present DroidOL, an accurate, adaptive and scalable Android malware detection framework.
DroidOL's unique feature is its ability to handle \textit{population drift} in Android malware through the use of online learning. DroidOL exhibits high accuracy, as it extracts effective structural features from apps using a state-of-the-art graph kernel. Further, DroidOL adapts automatically to the drift in malware characteristics over time and exhibits high scalability, making it suitable for real-world malware detection. 

Our large-scale evaluations on a real-world dataset demonstrates that DroidOL outperforms state-of-the-art malware detectors. DroidOL achieves 84.29\% accuracy outperforming existing techniques by more than 20\% in their typical batch-learning setting. This superior performance make DroidOL, in particular, and online learning based solutions, in general, better candidates for practical large-scale malware detection.

\section*{Acknowledgment}

We thank the authors of \cite{Drebin} and \cite{CSBD}, for their suggestions and discussions that helped us re-implement their methods. We thank Kevin Allix for sharing the dataset used in \cite{KevinThesis}.

\end{document}